# Optimizing Digital Adjudication through Social Network Analysis: An Empirical Study of Credit Card Disputes in Beijing

Authors: Chung Han Tsai(1,2)*, ChengTo Lin(2)**,   Baowen Zhang (2), Qingyue Deng(3), Yunhui Zhao(3), Zhijia Song(3)

1 School of Business Administration, Guangzhou Institute of Science and Technology，Guanzhou, China

2 Renmin University of China, Beijing, China

3 ShanDong University , Shandong, China

*Correspondence to: Chung-Han Tsai , Email：caizonghan@bucea.edu.cn

**Co-First Auhor:Cheng-To Lin

**Abstract**

Amid the rapid digitalization of judicial systems, the integration of big data into adjudication remains underexplored, particularly in uncovering the structural logic of legal applications. This study bridges this gap by employing social network analysis (SNA) to examine credit card disputes involving personal information protection adjudicated in Beijing (2022–2024). By constructing a legal citation network, we reveal the latent patterns of substantive and procedural law application. The findings demonstrate that SNA can effectively identify core legal norms and typify cases, offering a robust methodological framework for optimizing 'Digital Court' systems. These insights provide practical pathways for enhancing judicial efficiency and consistency through data-driven case retrieval and holistic judicial information networks.

1. Introduction

The Chinese judiciary is currently grappling with a critical structural imbalance: a surging caseload juxtaposed with limited judicial resources. As noted by Su (2010), the growth in litigation has consistently outpaced the expansion of judicial personnel, exacerbating the long-standing dilemma of "too many cases and too few judges". Empirical data highlight the severity of this "judicial overload": in 2022, judges in primary people's courts handled an average of 274 cases per capita, with figures exceeding 400 in certain jurisdictions (Cheng, 2022). Under these constrained circumstances, enhancing adjudicative efficiency without compromising substantive fairness has emerged as a central imperative for judicial reform.

In response to these dual pressures, China's courts have aggressively pursued digital transformation as a strategic institutional response (General Office of the CPC & State Council, 2023). This "Digital Court" initiative aims to optimize litigation workflows and alleviate judicial workloads through big data analytics (Jia, 2024). High-level mandates, such as those emphasized by Zhang Jun, President of the Supreme People's Court, envision a transition from traditional experience-based adjudication to a data-driven, scientifically informed paradigm (Zhang J., 2024). Pioneering jurisdictions like Beijing and Shanghai have already integrated online case handling and digital frameworks into their daily operations, demonstrating the preliminary effectiveness of these technical interventions (Zhao, 2024; Chen et al., 2024).

However, the current trajectory of digital court construction faces a critical bottleneck. While existing smart trial systems—categorized primarily into auxiliary tools and procedural supervision mechanisms—have succeeded in automating administrative tasks (Chen et al., 2024; Fan et al., 2024), they often lack the capacity to provide deep, substantive intellectual support for legal reasoning. Current academic research and practical applications largely rely on surface-level "keyword-matching" algorithms or theoretical abstractions (Wang, 2024; Chen & Sun, 2023). Consequently, the potential for big data to resolve deeper issues of trial quality control and adjudicative logic remains significantly underutilized due to a technology-centered approach that overlooks doctrinal coherence (Xu & Zhu, 2020).

To bridge this gap, this study introduces Social Network Analysis (SNA) as a novel methodological framework for quantitative legal research. Originating in sociology (Wellman, 1979), SNA is uniquely positioned to map the structural relationships between legal provisions. Unlike traditional bibliometrics, it reveals the latent logic of "analogy–induction–deduction" inherent in judicial decision-making . Recent studies have validated this approach in modeling criminal trial outcomes and statutory evolution, proving its efficacy in revealing interconnections among legal norms (Masías et al., 2016; Coupette et al., 2021) . By treating cited laws as nodes and their co-occurrence as ties, this approach transforms isolated judicial decisions into an interconnected judicial information network.

Focusing on credit card disputes involving personal information protection adjudicated in Beijing between 2022 and 2024, this paper constructs an empirical legal citation network to decode the patterns of substantive and procedural law application. This study contributes to the literature by demonstrating how data-driven network models can facilitate precise case typology, support intelligent legal retrieval, and ultimately promote judicial justice through enhanced consistency and efficiency .

2.Literature Review & Theoretical Framework
2.1 The Evolution and Limitations of Intelligent Trial Systems

The integration of artificial intelligence into the judiciary has precipitated a paradigm shift in case management. Existing judicial practices indicate that smart trial systems generally manifest in three distinct operational modalities. The first category comprises specific auxiliary adjudication tools, designed to extract key dispute elements—such as the "Shareholder's Right to Know" models developed in Shanghai—thereby enabling judges to rapidly identify evidentiary focal points (Chen et al., 2024) . The second category involves comprehensive process assistance, exemplified by the intelligent systems in Beijing's Xicheng District Court, which automate case assignment, file linkage, and enforcement procedures through big data integration (Fan et al., 2024) . The third category focuses on procedural supervision, utilizing automated algorithms to flag irregularities such as delayed lifting of enforcement measures or defects in service of process (Jia, 2024) .While these applications have significantly optimized procedural efficiency, academic discourse on intelligent adjudication remains disproportionately focused on theoretical abstractions rather than empirical validation. Crucially, a significant gap exists in the current landscape: existing systems prioritize procedural automation over the substantive analysis of adjudicative logic. Consequently, the "black box" of how judges actually select and combine legal provisions in practice remains largely opaque to existing digital systems (Wang, 2024; Xu & Zhu, 2020).

While these applications have significantly optimized procedural efficiency, academic discourse on intelligent adjudication remains disproportionately focused on theoretical abstractions rather than empirical validation. For instance, Wang (2024) proposes a conceptual framework for digitizing legal norms by mapping factual elements to legal facts, while Chen and Sun (2023) explore the use of pre-trained language models for legal text extraction . Although these studies offer valuable theoretical insights, they largely remain at the conceptual level, lacking systematic verification against large-scale judicial datasets.

Crucially, a significant gap exists in the current landscape: existing systems prioritize procedural automation and supervision over the substantive analysis of adjudicative logic. Legal reasoning in civil law jurisdictions, including China, typically follows an "analogy–induction–deduction" trajectory, where judges identify analogous precedents to induce general rules before applying them deductively to pending cases . However, current smart court technologies and related academic research rarely employ auxiliary models capable of systematically revealing these patterns of legal application. Consequently, the "black box" of how judges actually select and combine legal provisions in practice remains largely opaque to existing digital systems.

## 2.2 The Relational Structure of Credit Card Disputes

To apply computational methods effectively, one must first delineate the doctrinal structure of the disputes in question. Credit card cases involving personal information protection are not merely bilateral debt disputes; they involve a complex tripartite legal relationship among the Card Issuer (Bank), the Cardholder, and External Collection Agencies.

Scholars such as Xu (2023) and Chen (2024) have categorized these legal relationships, but often overlook the specific allocation of rights regarding personal data . As illustrated in Figure 1   the allocation of rights and obligations in these disputes is structurally distinct:

(i)The Primary Contractual Relationship: Exists between the Bank and the Cardholder, governed by the credit card service contract, focusing on the obligation to repay principal and interest.

(ii)The Entrustment Relationship: Exists between the Bank and the External Collection Agency, where the bank delegates collection rights while retaining supervisory duties.

(iii)The Tort/Privacy Interface: Although no direct contract exists between the Collection Agency and the Cardholder, the agency bears a statutory obligation to protect the cardholder's personal information during the collection process.

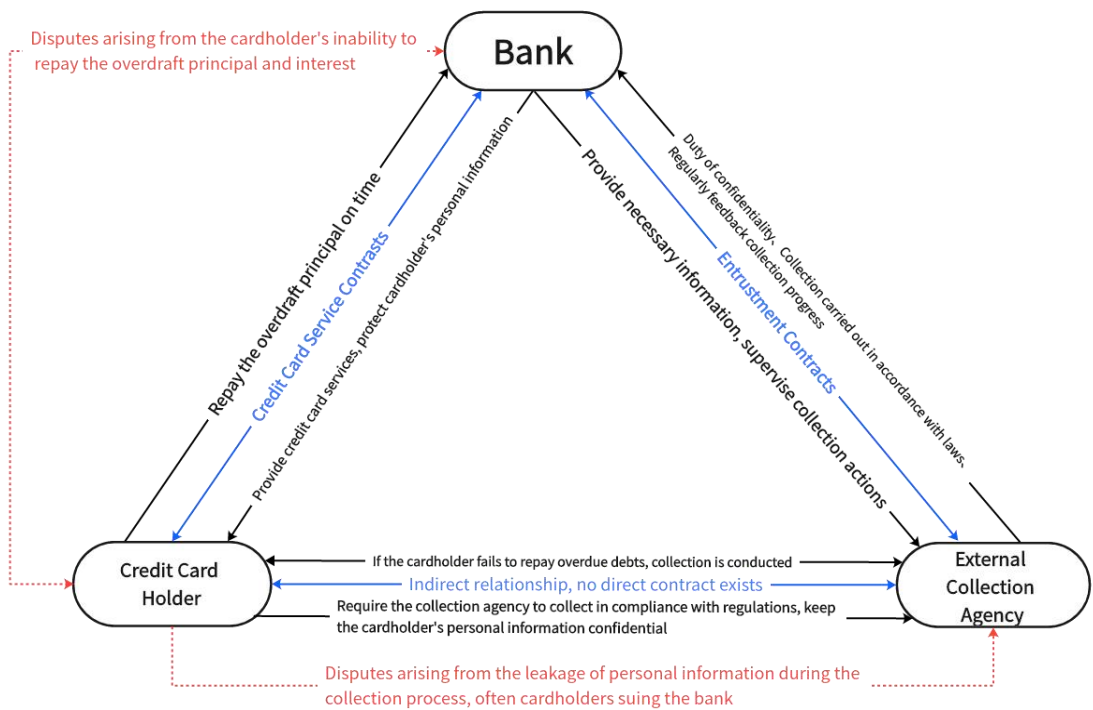

Figure1

(The diagram maps the contractual link between Bank and Cardholder, the entrustment link between Bank and Agency, and the indirect privacy obligations during debt collection)

This structured relational model provides the theoretical basis for our empirical analysis. It confirms that the legal issues in such cases are highly typified, making them ideal candidates for network-based modeling.

2.3 Social Network Analysis as a Legal-Empirical Framework

Building on this relational understanding, this study adopts Social Network Analysis (SNA) as its primary methodological framework. Originating in sociology to study the structural patterns of interaction among actors (Wellman, 1979), SNA has evolved into a versatile tool for uncovering latent relationships in complex systems.

The application of SNA to legal scholarship—often termed "Computational Legal Studies"—offers a robust alternative to traditional textual analysis. While standard bibliometrics or word-frequency models capture isolated data points, they fail to represent the relational structure essential to legal reasoning. Law is inherently relational; statutes and regulations do not operate in a vacuum but function as a dynamic network of interlinked norms. Recent scholarship has validated the efficacy of this approach: Masías et al. (2016) demonstrated that SNA offers superior explanatory power in modeling criminal trial outcomes, while Coupette et al. (2021) successfully utilized network analysis to map the evolution of statutory systems .

In the context of this study, SNA provides a unique lens to deconstruct the "judicial logic" of credit card disputes. By conceptualizing the legal provisions governing the relationships shown in Figure 1 as nodes and their co-application within a judgment as ties, we can construct a legal citation network. This structural approach allows for the visualization of how distinct legal norms are woven together in adjudication, revealing the "centrality" of specific laws and the "density" of their application.

## 3. Methodology & Data

### 3.1 Data Collection and Sampling Procedure

This study relies on judicial decisions retrieved from the Peking University Legal Expert Database (PKULaw), a comprehensive repository of Chinese legal documents. To ensure the representativeness of the sample regarding "Digital Court" practices, the scope was restricted to civil rulings adjudicated by courts in Beijing, spanning the period from January 1, 2022, to September 30, 2024.

### 3.2 Construction of the Legal Citation Network

The retrieval process employed a keyword combination of "credit card disputes" and "personal information protection." The initial search identified 49 judicial decisions. A systematic screening procedure was then applied:

(i) Deduplication: One duplicate record was removed.

(ii) Relevance Verification: Cases were manually reviewed to ensure they conformed to the standard typology of consumer credit disputes. For instance, a quadrilateral cluster identified in the preliminary network (labeled QMDH) was found to correspond to a specific case (Agricultural Bank of China Beijing Shunyi Branch v. Jin Yong) with a fundamentally different cause of action. To preserve the internal consistency of the dataset, this outlier was excluded.

The final valid sample comprised 48 unique judicial decisions, which served as the empirical basis for the subsequent social network analysis.

### 3.2 Construction of the Legal Citation Network

To quantify the "judicial logic," this study utilized Social Network Analysis (SNA) to model the relationships between cited legal provisions. The construction of the network followed a two-step mathematical transformation:

Step 1: Matrix Generation

An affiliation matrix (2-mode network) was established with legal provisions as rows and the 48 judicial decisions as columns. Let $X_{ij}$ denote the citation status:

$$X_{ij} = \begin{cases} 1 & \text{if provision is cited in judgment } j \\ 0 & \text{otherwise} \end{cases}$$

This "legal provisions–judicial decisions" matrix captures the raw co-occurrence data.

Step 2: Network Visualization

Using UCINET software, the affiliation matrix was projected into a 1-mode network $G(N, K)$. In this network, nodes (N) represent the legal provisions, and weighted ties (K) represent the frequency of their co-occurrence within the same judgments6. The visualized structure is presented in Figure 2.

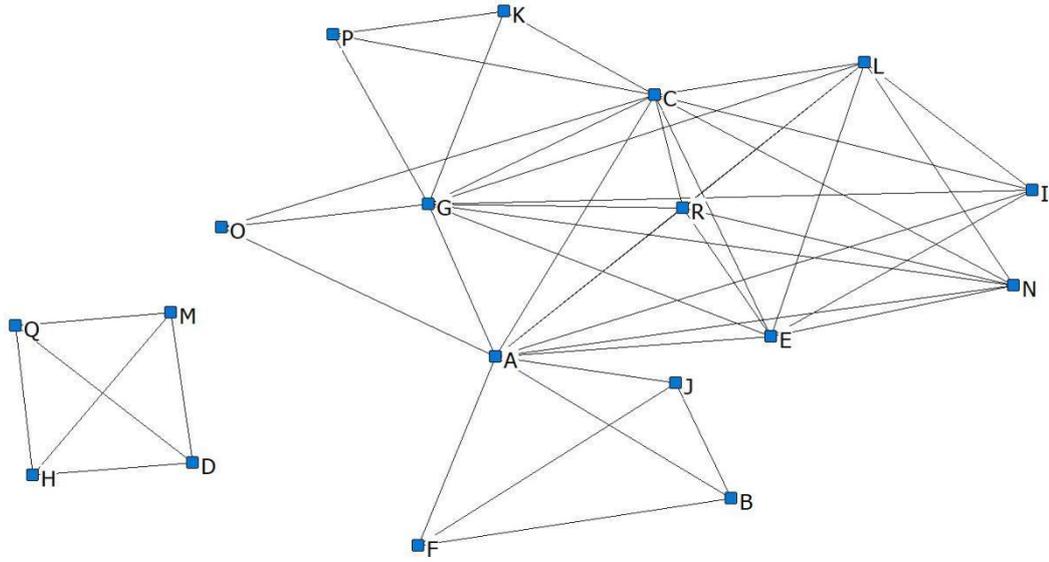

**Figure 2**
(The diagram illustrates the connectivity of legal norms, excluding the isolated QMDH cluster.)

This visualization (Figure 2) immediately reveals the structural characteristics of the case type, distinguishing between core legal clusters and peripheral citations .

3.3 Operationalization of Network Metrics

To interpret the structural properties of the network in Figure 2, we calculated indicators at both the individual node level and the overall network level.

1. Individual-Level Metrics

(i) Degree Centrality: Measures the number of direct connections a node has. In this context, a higher degree indicates a legal provision that is "frequently cited" and occupies a central position in adjudicative practice.

(ii) Betweenness Centrality ($C_B$): Measures the extent to which a node acts as a bridge along the shortest paths between other nodes. A high betweenness value suggests the provision plays a key role in coordinating different legal norms 9. The formula is:

$$C_B(n_i) = \frac{\sum_{j<k} g_{jk}(n_i)}{g_{jk}}$$

Where $g_{jk}$ represents the number of shortest paths between nodes $j$ and $k$, and $g_{jk}(n_i)$ denotes the number of those paths passing through node $i$.

2. Network-Level Metrics

(i) Network Size: Defined as the total number of nodes (legal provisions) in the network, reflecting the diversity of legal norms applied .

(ii) Network Density $D$: Measures the overall connectedness of the network, calculated as the ratio of actual edges to possible edges. A higher density indicates a tighter integration of legal application . The formula is:

$$D=\frac{2L}{g(g-1)}$$

Where L denotes the number of existing edges and g represents the total number of nodes .

By integrating these metrics, we can strictly evaluate whether the judicial application in credit card disputes exhibits the "clustering" and "consistency" required for digital court algorithms.

4. Results
4.1 Centrality Analysis: The Hierarchy of Legal Norms

The calculation of individual-level metrics—specifically degree centrality and betweenness centrality—reveals a distinct hierarchical structure in the application of law. As presented in Table 1, specific legal provisions exhibit significant structural advantages, functioning as the core "judicial anchors" for credit card disputes involving personal information.

Table 1: Degree and Betweenness Centrality of Nodes Related to Laws and Regulations  (Note: Nodes are labeled alphabetically corresponding to the network visualization)

| Legal Provisions | Node Label | Degree | Betweenness Centrality |
|---|---|---|---|
| *Interpretation on Retroactivity of Civil Code*, Art. 1 | A | 11 | 32.067 |
| *Civil Code*, Art. 1032 | B | 3 | 0 |
| *Contract Law*, Art. 60 (Invalidated) | C | 10 | 12.067 |
| *Civil Code*, Art. 496 | D | 3 | 0 |
| *Civil Code*, Art. 6 | E | 7 | 0.4 |
| *Civil Code*, Art. 1034 | F | 3 | 0 |
| *Contract Law*, Art. 107 (Invalidated) | G | 10 | 12.067 |
| *Civil Code*, Art. 497 | H | 3 | 0 |
| *Contract Law*, Art. 8 (Invalidated) | I | 5 | 0 |
| *Civil Code*, Art. 1035 | J | 3 | 0 |
| *Guarantee Law*, Art. 18 (Invalidated) | K | 3 | 0 |
| *Provisions on Civil Disputes over Bank Cards*, Art. 2 | L | 7 | 0.4 |
| *Measures for the Administration of Bank Card Business*, Art. 6 | M | 3 | 0 |
| *Civil Procedure Law*, Art. 147 | N | 6 | 0 |
| *Civil Procedure Law*, Art. 144 | O | 3 | 0 |
| *Interpretation on Retroactivity of Civil Code*, Art. 20 | P | 3 | 0 |
| *Measures for Supervision of Credit Card Business*, Art. 39 | Q | 3 | 0 |
| *Civil Procedure Law*, Art. 95 | R | 6 | 0 |

(Compiled and Illustrated by the Author)

**Analysis of Key Findings:**

First, Provisions on the Temporal Applicability of the *Civil Code* (Node A) occupy the most prominent position within the network, exhibiting the highest degree (11) and betweenness centrality (32.067). This high betweenness indicates that Node A serves as the critical "bridge" connecting various legal clusters. This phenomenon is explained by the specific temporal context of the sampled cases: many disputes involved personalized installment agreements (permitted under Article 70 of the Measures for Supervision of Credit Card Business) that extended across the enactment of the *Civil Code* in 2021 . Consequently, judges were required to navigate inter-temporal conflicts of law, making the retroactivity provision the structural pivot of legal reasoning .

Second, General Contract Principles (Nodes C and G) maintain a dominant role despite statutory succession. Articles 60 and 107 of the former Contract Law show high degree and betweenness centrality (Degree: 10; Betweenness: 12.067). Although the *Civil Code* has formally replaced these statutes, their continued citation reflects the stability of the underlying "rights-obligations" structure in credit card disputes—specifically, the primary obligation to repay principal and the ancillary obligation to protect data .

Third, Specialized Judicial Interpretations (Node L) appear less central than general principles. The Provisions on Civil Disputes over Bank Cards exhibit lower centrality values compared to the general contract provisions. This suggests that in disputes involving personal information, judicial practice relies more heavily on foundational contract law theories (performance and breach) rather than the specific regulations devised for internet finance risks .

4.2 Network Structure and Adjudicative Homogeneity

At the macro level, the structural characteristics of the legal citation network provide empirical evidence regarding the consistency of judicial logic. The overall network metrics are summarized in Table 2.

Table 2 Overall Metrics

| Metric | Value |
|---|---|
| **Density** | 0.301 |
| **Number of Edges** | 92 |
| **Number of Nodes** | 18 |

(Compiled and Illustrated by the Author)

**Interpretation of Network Density:** The network density of 0.301 exceeds the standard threshold for sparse networks (typically 0–0.25), indicating a relatively dense web of connections among legal provisions . This high density implies that judges consistently co-cite a specific cluster of legal norms when adjudicating credit card disputes involving personal information.

From a "Digital Court" perspective, this structural tightness confirms that this category of disputes is highly "typified" or homogeneous . The strong interconnections suggest a shared, stable adjudicative logic across different judges and cases. This finding empirically validates the feasibility of developing standardized, batch-processing algorithms for such disputes, as the legal application follows a predictable and recurring pattern .

Furthermore, the network analysis successfully identified structural outliers. As noted in the methodology, a disconnected quadrilateral cluster (Nodes Q, M, D, H) was detected and confirmed to represent a case with a fundamentally different cause of action. This demonstrates the network model's practical utility in automated case classification and "noise" filtering for judicial databases .

## 5. Discussion

### 5.1 Beyond the "Vending Machine": From Mechanical Application to Holistic Support

Max Weber famously echoed Montesquieu's metaphor of the judge as a "legal vending machine," where facts are input and judgments are mechanically output based on statutory codes. While this ideal underscores the normative goal of minimizing arbitrary discretion, the reality of the Chinese judiciary—characterized by complex legislative layers and "judicial overload"—renders such mechanical uniformity difficult to achieve .

Current "Smart Court" initiatives often attempt to replicate this mechanical model through automation. However, as our analysis suggests, adjudication is not a linear computation but a structured networking of legal norms. Therefore, we propose reconstructing the intelligent trial system not as a substitute "robot judge," but as a "Digital Holistic System". This system leverages big data to assist human judges by systematically processing the relational structure of cases, thereby supporting consistent reasoning while preserving the human element of justice .

The proposed framework for this reconstruction is illustrated in Figure 3 .

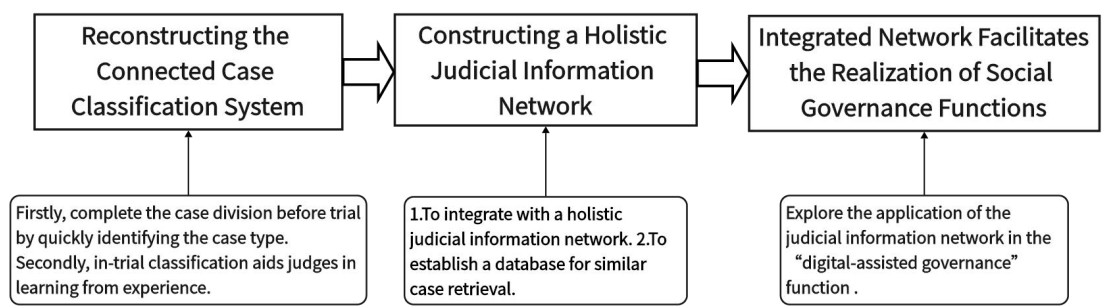

Figure 3

(The diagram outlines the three-stage pathway: Reconstructing Case Classification →Constructing a Holistic Judicial Information Network → Realizing Social Governance Functions.)

### 5.2 Reconstructing the Case Classification System

The first step in this holistic framework is to optimize how cases are identified and sorted.

1. Pre-trial: Network-Based Typology Identification Traditional judicial databases rely on keyword retrieval or formal causes of action, which are often too coarse to capture substantive legal issues . Our empirical results demonstrate that SNA can identify case types based on their structural position within a legal citation network. For instance, the isolation of the "quadrilateral cluster" (Figure 2) in our study allowed for the rapid detection of an atypical case that required separate processing . By integrating this logic, digital courts can automatically filter "batch cases" (high network density) from "complex cases" (structural outliers) at the filing stage, significantly streamlining pre-trial procedures .

2. In-trial: Experience-Informed Learning Current AI classification remains heavily dependent on surface-level rules. A network-based system, however, can incorporate "judicial experience" by learning from the exclusion and clustering decisions made by expert judges. This "human-in-the-loop" approach allows the system to evolve from mechanical categorization to a refined, experience-based classification model, offering judges reference cases that are substantively, not just nominally, similar .

5.3 Constructing a Holistic Judicial Information Network

The second step is to integrate fragmented judicial data into a connected ecosystem centered on res judicata.

1. Visualizing the "Chain of Justice" Judicial decisions are often treated as isolated data points. By constructing a Holistic Judicial Information Network, these points are transformed into a visualized structure . This network allows judges to observe the "core path" of legal application for specific dispute types. For credit card disputes, the strong linkage between Contract Law Art. 60 and Art. 107 (as identified in Table 1) serves as a normative baseline. Deviations from this baseline can trigger automated alerts, enabling courts to monitor adjudicative quality and reduce discretionary inconsistencies .

2. A Networked Res Judicata System China's current procedural framework lacks a systematic mechanism for coordinating the binding effect of res judicata across different court levels . A network-based retrieval system addresses this by integrating typified cases, judicial interpretations, and guiding cases into a unified reference structure. This constrains discretion through authoritative precedents, enhancing the predictability and certainty of legal outcomes .

3. Dynamic Legal Monitoring The network also functions as a monitor for legal implementation. Our finding that the "Retroactivity Provision" (Node A) occupies the highest centrality reflects a specific transitional tension in the law . Digital courts can utilize such network centrality metrics to identify "hotspots" of legal conflict, providing real-time feedback to legislators and the Supreme People's Court regarding the interaction between new codes and legacy contracts.

5.4 Ethical Boundaries and Potential Risks

While the efficiency gains of such a system are evident, the pursuit of "algorithmic justice" entails significant risks .

(i) Technical Risks: Limitations in data quality or algorithmic design may lead to inaccurate recommendations or "algorithmic bias," potentially entrenching erroneous precedents if not critically reviewed.

(ii) Ethical Accountability: The allocation of responsibility for errors in AI-assisted adjudication remains a contested issue. If a judge relies on an algorithmic recommendation that proves incorrect, the question of accountability—whether it lies with the judge, the system designer, or the institution—is complex .

Therefore, we maintain that artificial intelligence must remain strictly auxiliary.

The authority to render final judgments and the duty to assess the specific context of each case must reside with human judges . The construction of digital courts must proceed with caution, ensuring that technological innovation serves to augment, rather than erode, the human-centered nature of justice.

6.Conclusion

6.1 Summary of Research Findings

This study empirically examined the adjudicative patterns of credit card disputes involving personal information protection, utilizing a dataset of 48 validated judicial decisions adjudicated in Beijing between 2022 and 2024. By constructing a database of cited laws and applying Social Network Analysis (SNA), we successfully mapped the latent structure of legal application in this specific domain.

The quantitative results reveal two critical insights. First, the high network density ($D$=0.301) indicates that the adjudicative logic for these disputes is highly homogenized, providing strong empirical support for the implementation of batch-processing algorithms in digital courts. Second, the centrality analysis identified specific legal provisions—most notably the transitional rules regarding the *Civil Code* and foundational contract principles—as the structural "anchors" of judicial reasoning. These findings demonstrate that "judicial logic" is not an abstract concept but a quantifiable network structure

6.2 Theoretical and Practical Implications.

Theoretically, this research contributes to the field of "Computational Legal Studies" by validating SNA as a robust framework for decoding legal complexity. It transcends traditional keyword-based retrieval by offering a relational perspective on how judges navigate normative systems.Practically, the study proposes a three-step implementation pathway for optimizing "Smart Trial Systems" :

(i) Case Clustering: Moving from formal cause-of-action classification to network-topology-based identification, enabling the automated separation of routine and complex cases.Network Linking: Constructing a holistic judicial information network that connects isolated judgments, thereby enhancing the visibility of res judicata and supporting adjudicative consistency.

(ii) Service Enhancement: Utilizing the network for dynamic legal monitoring and "data-assisted governance," shifting the digital court function from passive archiving to proactive intellectual support.By aligning technological tools with the doctrinal needs of legal reasoning, this framework offers a practical blueprint for advancing judicial efficiency while safeguarding the substantive quality of justice.

6.3 Limitations

Despite the rigorous methodology, this study is subject to several limitations that warrant consideration:

> (i) Geographic Scope: The empirical analysis was restricted to cases adjudicated in Beijing. While Beijing represents a leading jurisdiction in digital reform, the extent to which the identified legal application patterns are replicable in other regions with different economic and judicial conditions remains to be verified .
> 
> (ii)Data Source Constraints: The study relied exclusively on the PKU Law

Database. Although authoritative, reliance on a single source may introduce potential selection bias regarding case completeness compared to internal court systems .

(iii) Network Dimensionality: The current analysis focused solely on the citation network of laws and regulations. It did not incorporate other potential network dimensions, such as factual elements, procedural stages, or litigant relationships, which could offer complementary insights into case similarity .

6.4 Future Research Directions

Building on these findings, future research should aim to expand the boundaries of the judicial data network8.Expansion of Scope: Future studies should incorporate multi-regional datasets to test the generalizability of the network models across different jurisdictions.Complex Case Modeling: Research should move beyond simple batch cases to explore how SNA can deconstruct complex disputes into combinations of simpler "case modules"9.Multi-Dimensional Integration: Future models should attempt to construct integrated networks that link cases not just by legal citations, but also by temporal, geographic, and procedural keywords. Such "multi-layer" networks would further enhance the capacity of digital courts to assist judges in experience summarization and decision-making, ultimately contributing to the development of a fully holistic judicial information system .

Data availability

To assess the reliability and internal consistency of the research data, this study employs Cronbach's alpha (α) coefficient and conducts the analysis using SPSSAU software. The results are reported in Table 4.

| Name | Corrected Item-Total Correlation (CITC) | Cronbach's a ifItem Deleted | Cronbach's α |
|---|---|---|---|
| Item 1 | 0.758 | - | 0.586 |
| Item 2 | 0.758 | - | 0.586 |
|  | Note:Standaerdized Cronbach's α=0.863 |  |  |

Table 4 Cronbach's Alpha Reliability Analysis Results

As shown in Table 4, the Cronbach's α coefficient for the BA, PC, and SN variables is 0.586, which exceeds the commonly accepted minimum threshold of 0.5 but remains below 0.6. It should be noted that this scale consists of only two measurement items, a condition under which Cronbach's α values are often relatively conservative.

In addition, the corrected item–total correlation (CITC) values for all measurement items exceed 0.5, indicating strong correlations between individual items and the overall scale. This suggests an acceptable level of internal consistency among the items. Taken together, these results indicate that the research data demonstrate an adequate level of reliability and are suitable for subsequent empirical analysis.

Competing interests

The author(s) declare no competing interests.

Ethical statements

This article does not contain any studies with human participants performed by any of the authors.

**Numbered Endnotes**